\documentclass[a4paper,12pt]{article}

\usepackage{amssymb}%,latexsym} 

\usepackage[utf8]{inputenc}

\usepackage{fullpage}

\usepackage{boxedminipage}

\usepackage{listings}

\usepackage{minitoc}

\makeatletter \@addtoreset{equation}{section} 
\makeatother

\newif\ifpdf \ifx\pdfoutput\undefined \pdffalse 

\else \pdfoutput=1 

\pdftrue \fi

\ifpdf 
\usepackage[pdftex]{graphicx} \else 
\usepackage{graphicx} \fi 
\begin{document}

\ifpdf \DeclareGraphicsExtensions{.pdf, .jpg, .tif} \else \DeclareGraphicsExtensions{.eps, .jpg} \fi 
\begin{titlepage}

\thispagestyle{empty} 
\begin{flushright}
	\hfill{CERN-PH-TH/2006-016} \\
	\hfill{hep-th/0602045} 
\end{flushright}

\vspace{35pt} 
\begin{center}
	{ \LARGE{\bf Non-K\"ahler attracting manifolds }}
	
	\vspace{60pt}
	
	{\bf Gianguido Dall'Agata }
	
	\vspace{30pt}
	
	{\it Physics Department,\\
	Theory Unit, CERN, \\
	CH 1211, Geneva 23, \\
	Switzerland}
	
	\vspace{15pt}
	
	\vspace{40pt}
	
	{ABSTRACT} 
\end{center}

\vspace{10pt} 

We observe that the \textit{new attractor} mechanism describing IIB flux vacua for Calabi--Yau compactifications has a possible extension to the landscape of non-K\"ahler vacua that emerge in heterotic compactifications with fluxes.
We focus on the effective theories coming from compactifications on generalized half-flat manifolds, showing that the Minkowski ``attractor points'' for 3-form fluxes are special-hermitian manifolds.

\end{titlepage}
\newpage \baselineskip 6 mm

\tableofcontents

\bigskip

\section{Introduction}

The black hole attractor mechanism \cite{Ferrara:1995ih,Ferrara:1997tw} is a translation of the no-hair theorem into algebraic equations that specify the values of the moduli at the horizon, in terms of the black hole mass and charges.
These, in turn, specify the area of the horizon and hence the black hole entropy $S$.
For extremal black holes, the latter is controlled by the extrema of a potential $V_{BH}$ depending on the black hole central charges ${\cal Z}^{IJ}$: $S = \pi V_{BH}|_{hor}$ \cite{Andrianopoli:1996ve}.
In ${\cal N} = 2$ supergravity, thanks to the special-K\"ahler structure of the vector-multiplets moduli space, the central charge ${\cal Z}$ is covariantly holomorphic ($\bar D_{\bar \imath} {\cal Z} =\left( \bar 
\partial_{\bar \imath}- \frac12 K_{\bar \imath}\right){\cal Z}=0$) and the black hole potential is \cite{Ferrara:1996dd,Ceresole:1995ca} 
\begin{equation}
V_{BH} = |{\cal Z}|^2 + |D_i {\cal Z}|^2.
\label{VBH} 
\end{equation}
The supersymmetric configurations are then specified by the minimization condition 
\begin{equation}
D_i {\cal Z} = \left( 
\partial_i + \frac12 K_i\right){\cal Z} = 0 \label{DZ} 
\end{equation}
of the central charge, the latter being given by the invariant constructed from the symplectic vector of charges ${\cal Q} = (p^{\Lambda}, q_{\Lambda})$ and the symplectic sections of the moduli space ${\cal V} = (L^{\Lambda}, M_{\Lambda})$ \cite{Ceresole:1995jg} 
\begin{equation}
{\cal Z}= \langle {\cal Q}, {\cal V}\rangle = q_{\Lambda} L^\Lambda- p^{\Lambda} M_{\Lambda}.
\label{potbh} 
\end{equation}
The supersymmetric attractor equation is the algebraic translation of (\ref{DZ}) \cite{Ferrara:1995ih,Strominger:1996kf,Ferrara:1996dd}:
\begin{equation}
{\cal Q} = 2 {\rm Re }(-i \bar {\cal V} {\cal Z}).
\label{susyattbh} 
\end{equation}
The asymptotically flat 4-dimensional ${\cal N} = 2$ black holes arise from string theory compactifications on Calabi--Yau manifolds in the presence of non-trivial 5-form fluxes on the internal 3-cycles.
At the horizon, the 10-dimensional space is the product AdS$_2 \times S^2 \times CY_{pq}$, and the attractor equation can be understood as the expansion of the integral 5-form fluxes in terms of the holomorphic cycles of the 3-form cohomology \cite{Ceresole:1995ca,Moore:1998pn} 
\begin{equation}
F_5 = 2 {\rm Re}({\cal Z} \widehat \Omega) \wedge \omega_{S^2}, \label{5formattr} 
\end{equation}
where $\omega_{S^2}$ is the volume form for $S^2$, and $\widehat \Omega$ is the normalized holomorphic form of $CY_{pq}$, which is an ``attractive variety'' \cite{Moore:1998pn}.
The same central charge ${\cal Z}$ can be expressed in terms of the Calabi--Yau data as the integral of the fluxes 
\begin{equation}
{\cal Z} = \int_{CY\times S^2} F_5 \wedge \widehat \Omega.
\label{ZCY} 
\end{equation}

This form of the attractor equations has a striking similarity with the equations that specify the vacua of IIB string theory Calabi--Yau compactifications to 4 dimensions, in the presence of 3-form fluxes and O3/O7-planes.
The superpotential of the effective ${\cal N} = 1$ theory is a holomorphic function of the complex-structure moduli 
\begin{equation}
W = \int_{CY} G \wedge \Omega(z^i), \label{supoIIB} 
\end{equation}
where $\Omega$ is the holomorphic form and $G = F_{\rm RR} - \tau H_{\rm NS}$ is the complex 3-form flux, depending on the axion/dilaton $\tau$ and constructed from the Ramond--Ramond 3-form $F_{\rm RR}$ and the Neveu--Schwarz one $H_{\rm NS}$.
From the superpotential one can define the quantity $Z = {\rm e}^{K/2} W$, which is covariantly holomorphic and whose explicit form is the same as (\ref{potbh}), but with the flux charges $(p^\Lambda, q_{\Lambda})$ also depending on $\tau$, e.g.~$q_{\Lambda} = q_{\rm RR} - \tau q_{\rm NS}$.
The supersymmetric critical points are then given by 
\begin{equation}
D_i Z = \left( 
\partial_i + \frac12 K_i\right) Z = {\rm e}^{K/2}\left( 
\partial_i + K_i\right) W = {\rm e}^{K/2} D_i W = 0, \label{susycp} 
\end{equation}
which has the same form as (\ref{DZ}).
Also, the full potential of the effective theory can be expressed in terms of $Z$ as 
\begin{equation}
V=|D_i Z|^2 - 3 |Z|^2.
\label{potentialN1} 
\end{equation}
These similarities have recently led to the discovery of the \textit{``new attractor''} equations by Kallosh \cite{Kallosh:2005bj,Kallosh:2005ax}, using the special-K\"ahler structure of the moduli space inherited from the Calabi--Yau.

The new algebraic attractor equations can be expressed as a constraint coming from the reality of the fluxes, once the setup is uplifted to F-theory, in analogy with the black hole (\ref{5formattr}).
In this case the RR and NS 3-form fluxes merge in a real 4-form flux $F_4$, which has a non-trivial expectation value on a Calabi--Yau 4-fold $Y_{8}$.
The expansion of this 4-form is now performed with respect to the basis of 4-forms on $Y_{8}$ \cite{Kallosh:2005ax}: 
\begin{equation}
F_4 = 2 {\rm Re} \left[\bar Z \widehat \Omega_4 - \bar D^{\bar A}\bar Z D_{A} \widehat \Omega_4 + \bar D^{\tau \bar I}\bar Z D_{\tau I} \widehat \Omega_4 \right], \label{F4} 
\end{equation}
where $A = \{\tau,I\}$.
This expression is valid at any point in moduli space.
The attractor conditions are obtained by plugging the stationary-point conditions $DV = 0$ in (\ref{F4}).
For instance, the supersymmetric vacua are obtained by substituting $D_A Z = 0$ in (\ref{F4}), while the non-supersymmetric ones are obtained by using $2 D_i Z \, \bar Z = D_i D_j Z\, g^{j\bar \jmath} \bar D_{\bar \jmath} \bar Z $ (see also \cite{Kallosh:2006}).
Once again, we can rewrite this equation in an algebraic relation for the symplectic vector of the flux charges ${\cal Q} = (p_{\rm NS}^{\Lambda}, q^{\rm NS}_{\Lambda},p_{\rm RR}^{\Lambda}, q^{\rm RR}_{\Lambda})$, now doubled because of the appearance of the axion/dilaton $\tau$ \cite{Kallosh:2005ax}: 
\begin{equation}
{\cal Q} = \left( 
\begin{array}{c}
	2 {\rm Re}(Z \bar{\cal V}) \\[2mm]
	2 {\rm Re}(Z \bar \tau \bar{\cal V}) 
\end{array}
\right)+\left( 
\begin{array}{c}
	2 {\rm Re}(\bar Z^{0I} D_I {\cal V}) \\[2mm]
	2 {\rm Re}(\bar Z^{0I} \bar \tau D_I{\cal V}) 
\end{array}
\right),
\label{attrIIB} 
\end{equation}
where $Z^{0I}= D^{0I}Z$.
These equations not only describe both supersymmetric and non-supersymmetric attractors, but they can also describe Minkowski vacua for which $Z = 0$ \cite{Kallosh:2005ax}.
These points cannot be obtained by the black hole attractor equations (\ref{susyattbh}), because $Z = 0$ would mean a singular solution, with zero area of the horizon.

In the following we try to extend this new attractor mechanism to string theory compactifications on non-K\"ahler manifolds in the presence of fluxes.
Despite most of the literature on the lanscape of flux vacua considers only the IIB case on Calabi--Yau's, it is known that there is a huge part of this landscape that involves compactifications on non-K\"ahler manifolds.
This is actually the generic case that arises when considering the flux back-reaction.
This type of solutions, first found in \cite{Strominger:1986uh} for the common sector of string theory, also appear in type II when more general fluxes are considered.
It is also clear that T-duality transformations of backgrounds with fluxes give rise to new backgrounds that involve geometric deformations leading to non-K\"ahler manifolds \cite{Kachru:2002sk,Gurrieri:2002wz};  the corresponding effective theories show an interesting dependence of the potential on the size moduli, in addition to the complex-structure moduli dependence of ordinary Calabi--Yau reductions.

The drawback of considering such compactifications is the lack of a good description of the moduli space.
However, an important step forward was done in \cite{Grana:2005ny}, where the authors showed that, for a generic SU(3) structure manifold, the moduli space of the effective theory is still described by a special-K\"ahler manifold.
Using this result, we can work by analogy with \cite{Kallosh:2005bj,Kallosh:2005ax} to extend the new attractor equations.
We will argue that the conditions for critical points of the potential can be expressed once more as a relation between a charge vector ${\cal Q}$ (now including the ordinary flux charges and the geometric deformations, from now on named ``geometric fluxes''), the covariantly holomorphic superpotential (or central charge) $Z = {\rm e}^{K/2}W$, and the symplectic sections of the moduli space ${\cal V}$.
The covariantly holomorphic superpotential can still be described as a symplectic product 
\begin{equation}
Z = \langle {\cal Q} ,{\cal V}\rangle, \label{attraexample} 
\end{equation}
where the charges ${\cal Q}$ are now collected in a matrix, which is a double symplectic vector with respect to the complex structure and K\"ahler deformations, and ${\cal V}$ collects the symplectic sections of both parts of the moduli space.
The attractor equations can then be obtained using the reality of ${\cal Q}$ (or of the corresponding fluxes, both the ordinary ones and the geometric ones) and their expansion in terms of the basis of forms related to the light degrees of freedom of the effective theory.
For the common sector of string theory this reads 
\begin{equation}
\begin{array}{rcl}
	{\cal Q} &=& -2 {\rm Im}\left(\overline Z {\cal V} + g^{\alpha \bar \beta}\; {\rm e}^{-\widehat J_c} \otimes D_{\alpha}\widehat\Omega \; \overline D_{\bar \beta}\overline Z+ g^{i \bar \jmath}\;D_{i}{\rm e}^{-\widehat J_c} \otimes \widehat\Omega\; \overline D_{\bar \jmath}\overline Z \right.\\[2mm]
	&& \left.+ g^{\alpha \bar \beta} g^{i \bar \jmath}\; D_{i}{\rm e}^{-\widehat J_c} \otimes D_{\alpha}\widehat\Omega\;\overline D_{\bar \jmath}\overline D_{\bar \beta}\overline Z\right),
\end{array}
\label{attractorsummary} 
\end{equation}
where the greek and latin indices label the complex structure and K\"ahler moduli respectively.
Also here, as for the IIB case, this expansion is valid at any point in moduli space, but it becomes a non-trivial equation on the moduli/charges if one uses the condition that the potential (\ref{potentialN1}) is minimized $DV = 0$ (using $DZ = 0$ for supersymmetric vacua).

The plan of the paper is the following.
As a first step, in section \ref{sec:preliminaries} we discuss the moduli space of the effective theories on non-K\"ahler manifolds, introducing the generalized half-flat manifolds, which we will use in the following, and their differential algebra.
In section \ref{sec:attractors} we focus on the common sector of string theory and especially on heterotic compactifications deriving the new attractor equations for this case.
First we will deal with the case of a single size modulus, showing the analogy with the IIB case and then give general expressions of the attractor equations for the generic case.
In this same section we will show how the attractor points, for the case of vanishing 4-dimensional cosmological constant, lead to the restriction to special-hermitian manifolds.
Finally, in section \ref{sec:typeII} we comment on the extensions to the case of type II strings.

\section{Preliminaries}\label{sec:preliminaries}

In this section we introduce the necessary ingredients to describe the moduli space of non-K\"ahler compactifications.

Let us start with a lightning review of the elements related to the group structures of the tangent bundle of the compactification manifold, as these are very useful to discuss the non-K\"ahler backgrounds.
When the Calabi--Yau condition is relaxed, because of the presence of form fluxes, the compactifying manifold no longer has an SU(3) holonomy, but rather it shows an SU(3) structure.
This means that there are still an almost complex structure $J$ and a holomorphic form $\Omega$, which are globally defined, but they are not closed in general, i.e.~$dJ \neq 0$ and $d \Omega \neq 0$.
The SU(3) structures are classified by the ``intrinsic torsion'' $\tau$ that one has to add to the Levi-Civita connection $\nabla$, so that $\nabla(\tau) J = 0 = \nabla(\tau) \Omega$.
This torsion is completely determined by the exterior differentiation of the globally defined forms 
\begin{eqnarray}
d \widehat J & = & \frac34 i \left(W_1 \overline{\widehat \Omega} - \overline W_1 \widehat \Omega\right) + \widehat J \wedge W_4 + W_3 , \label{dJ} \\[2mm]
d \widehat \Omega & = & W_1 \widehat J \wedge\widehat J +\widehat J \wedge W_2 + \widehat \Omega \wedge \overline W_5 , \label{dOmega} 
\end{eqnarray}
where $\widehat J \wedge W_3 = \widehat \Omega \wedge W_3 = \widehat J \wedge \widehat J \wedge W_2 = 0$ and $\widehat J$ and $\widehat \Omega$ are the normalized versions $||\widehat \Omega|| = ||\widehat J || = 1$ of the globally defined forms $J$ and $\Omega$.
The Calabi--Yau condition is given by the vanishing of all the torsion classes $W_1 = \ldots = W_5 = 0$.
Since in the following we will be mainly concerned with the compactifications of the heterotic theory, we point out that the allowed torsion classes for the common sector is given by \cite{Cardoso:2002hd}
\begin{equation}
W_1 = W_2 = 0, \quad W_3 = 2 \star H^0, \quad 2 W_4 = - W_5 = 
\partial \Delta, \label{hettorsion} 
\end{equation}
where $H^0$ is the primitive part of the flux and $\Delta$ is the warp factor.
The latter is also proportional to the dilaton $d \Delta = d \phi$.

In \cite{Grana:2005ny,House:2005yc} it was shown that for compactifications on SU(3) structure manifolds $Y_6$, the space of metric deformations $\delta g_{mn}$ is related to the structure deformations $\delta J$ and $\delta \Omega$ in a fashion similar  to the one of Calabi--Yau compactifications \cite{Candelas:1990pi}.
In general, $\delta J$ and $\delta \Omega$ contain more degrees of freedom than the metric as the structure deformations parametrize $\frac{GL(6,{\mathbb R})}{SU(3)}$, while the metric deformations are elements of $\frac{GL(6,{\mathbb R})}{SO(6)}$.
However, these extra deformations can be removed by the use of local symmetries \cite{Grana:2005ny}.
In addition, the reduction to an ordinary 4-dimensional theory without massive gravitino multiplets implies the need for a truncation, where all the deformations that transform as triplets of SU(3) are removed \cite{Grana:2005ny}.
Once this is done, the moduli space of the complex structure moduli and of the K\"ahler ones (where the complex moduli are obtained by considering also the degrees of freedom from the NS 2-form $B$) parametrizes the special K\"ahler manifold 
\begin{equation}
{\cal M}_{TOT} = {\cal M}_{\Omega} \otimes {\cal M}_{J}, \label{spSu3} 
\end{equation}
with K\"ahler potentials given by 
\begin{eqnarray}
K_{J} & = & - \log i \langle {\rm e}^{- J_c} , {\rm e}^{- \overline J_c}\rangle = - \log \frac43 \int J \wedge J \wedge J \label{KJ} \\[2mm]
K_{\Omega} & = & - \log i \langle \Omega, \overline \Omega\rangle = - \log i \int \Omega \wedge \overline \Omega, \label{KOmega} 
\end{eqnarray}
where $J_c = B + i J$ and the brackets $\langle,\rangle$ denote the Mukai pairing:
\begin{eqnarray}
\langle \psi^{+}, \chi^{+}\rangle & = & \psi^+_0 \wedge \chi^+_6 - \psi^+_2 \wedge \chi^+_4 + \psi^+_4 \wedge \chi^+_2 - \psi^+_6 \wedge \chi^+_0, \label{Mukai1} \\[2mm]
\langle \psi^{-}, \chi^{-}\rangle & = & -\psi^-_1 \wedge \chi^-_5 + \psi^-_3 \wedge \chi^-_3 - \psi^-_5 \wedge \chi^-_1. \label{Mukai2} 
\end{eqnarray}
Here, the $\pm$ superscript refers to the grade of the $\psi^{\pm}$, $\chi^{\pm}$ components in $\Lambda^{\rm even|odd} T^* Y_6$.
From these relations we can infer that ${\rm e}^{- J_c}$ and $\Omega$ are the symplectic sections of the Hodge bundle of the moduli spaces of the K\"ahler and complex structure deformations respectively.
In the same way we can introduce covariantly holomorphic sections ${\cal V}$ corresponding to the normalized forms 
\begin{equation}
{\cal V} = ({\rm e}^{- \widehat J_c}\otimes \widehat \Omega) = {\rm e}^{(K_J + K_\Omega)/2}({\rm e}^{- J_c}\otimes \Omega).
\label{sec} 
\end{equation}
These sections obey 
\begin{equation}
\langle {\cal V}, \overline {\cal V}\rangle = - i, \label{normalization} 
\end{equation}
where the brackets are now the symplectic product of the sections or the Mukai pairing according to the representation of ${\cal V}$.

For Calabi--Yau compactifications the $T$ and $U$ moduli correspond to the expansion of ${\rm e}^{- J_c}\otimes \Omega$ on the basis of harmonic forms.
In detail: 
\begin{eqnarray}
{\rm e}^{- J_c} & = & X^0(T) + X^{i}(T) \omega_i - F_{i}(T) \widetilde \omega^i - F_0(T) \star 1, \label{Jcexp} \\[2mm]
\Omega & = & X^\Lambda(U) \alpha_{\Lambda} - F_{\Lambda}(U) \beta^{\Lambda}, \label{Omegaexp} 
\end{eqnarray}
where $(\alpha_{\Lambda},\beta^{\Lambda})$ are a base for the 3-forms and $1, \omega_i, \tilde \omega^{i}, \star 1$ are a base for the 0-,2-,4- and 6-forms.
Obviously in this case we can no longer use the harmonic forms as we are expecting that $dJ$ and/or $d \Omega$ may no longer be closed.

Following \cite{Grana:2005ny} we can still consider an expansion over a basis of 2-,3- and 4-forms that correspond to a truncation of the space of forms to a finite-dimensional subspace.
With some reasonable assumptions, the spectrum of such forms gets restricted to $\Lambda^{0,2,4,6}$, where $\Lambda^0$ is the constant function, $\Lambda^6$ is the volume form, and $\Lambda^2$, $\Lambda^4$ are spanned by the basis $\omega_i$, $\widetilde \omega^i$, with the same dimension $b_{\rm even}$, and $\Lambda^3$, of dimension $2(b_{\rm odd} + 1)$, is spanned by $(\alpha_{\Lambda}, \beta^{\Lambda})$ that form a symplectic set of basis forms.
These forms satisfy the completeness relations 
\begin{equation}
\langle \alpha_{\Lambda}, \beta^{\Sigma}\rangle= - \langle \beta^{\Sigma}, \alpha_{\Lambda}\rangle = \delta_{\Lambda}^{\Sigma}, \qquad \langle \alpha_{\Lambda}, \alpha_{\Sigma}\rangle = 0 = \langle \beta^{\Lambda}, \beta^{\Sigma}\rangle; \label{compl} 
\end{equation}
introducing $\omega_{I} = (1, \omega_{i})$, $\widetilde \omega^I = (\star 1, \tilde \omega^i)$ so that also ($\omega_{I}, \tilde \omega^I $) form a symplectic basis\footnote{Strictly speaking, the symplectic basis is really spanned by $(\lambda \omega_I, \mu \tilde \omega^{I})$, where $\lambda,\mu$ are anticommuting numbers \cite{Candelas:1990pi} and in the symplectic product one should also perform an integration over $d \lambda d \mu$.
However, using the Mukai pairing one gets the appropriate signs and relations, also using $\langle \omega_I, \omega_J\rangle = 0$.}, we obtain
\begin{equation}
\langle \omega_I, \tilde \omega^{J} \rangle = - \langle\tilde \omega^{J}, \omega_I \rangle= \delta_I^J.
\label{ome} 
\end{equation}
In \cite{Grana:2005ny} the further restriction that $\Lambda^1 = \Lambda^5 = 0$ was imposed, justified by the request that no SU(3) triplets appear in the reduction.
This is expected when the warp factor function is constant.
In this case, the supersymmetric backgrounds no longer contain such forms in the definition of the intrinsic torsion (\ref{hettorsion}).
However, on a more general ground, we can note that also in the case when the warp factor is non-trivial over the internal manifold, the moduli are always defined through $J$ and $\Omega$ and not through the normalised forms $\widehat J$ and $\widehat \Omega$.
For this reason we can always define a holomorphic form $\Omega = ||\Omega|| \widehat \Omega = {\rm e}^{\Delta} \widehat \Omega$, with $d \Omega = 0$ (following from $W_5 = 
\partial \log ||\Omega||$ in the heterotic case), or at least generically without triplets in the exterior differentiation.
The K\"ahler potential generated by $\Omega^{\prime} = {\rm e}^{\Delta} \Omega$ is $K_{\Omega^{\prime}} = K_{\Omega} - 2 \Delta$.
In the same way, a rescaled $J$ can be defined that does not give rise to triplets.
It is useful that, since $W_4$ and $W_5$ transform in the vector representation of SU(3) and we do not expect globally defined vector fields for this type of manifolds (otherwise the structure would be further reduced), we expect these torsion classes to be proportional to the derivative of the dilaton/axion and/or to the warp factor of the background.
We can therefore assume that the moduli space of the SU(3) structure manifolds that appear in the non-K\"ahler compactifications are given by spaces with $\Lambda$ spanned by $1,\omega_i, \alpha_{\Lambda}, \beta^{\Lambda}, \tilde \omega^i, \star 1$, where these forms satisfy the orthogonality relations\footnote{This translates to the condition $J \wedge \Omega = 0$ for any choice of $J$ and $\Omega$ in $\Lambda^2$ and $\Lambda^3$, respectively.} 
\begin{equation}
\omega_i \wedge \alpha_{\Lambda} = 0 = \omega_i \wedge \beta^{\Lambda}, \label{zero} 
\end{equation}
and the differential conditions \cite{deCarlos:2005kh,DAuria:2004tr,Grana:2005ny}
\begin{eqnarray}
d \omega_i & = & m_i^{\Lambda} \alpha_{\Lambda} - e_{i \Lambda} \beta^{\Lambda}, \label{domi} \\[2mm]
d \tilde \omega^i & = & 0, \label{domtil} \\[2mm]
d \alpha_{\Lambda} & = & e_{i \Lambda}\tilde \omega^i, \label{dalfa} \\[2mm]
d \beta^{\Lambda} & = & m_i^{\Lambda} \tilde \omega^i.
\label{dbeta} 
\end{eqnarray}
The charges $V_i = (m_i^{\Lambda},e_{i \Lambda})$ form a symplectic vector satisfying the constraint 
\begin{equation}
\langle V_i, V_j \rangle = 0, \label{const} 
\end{equation}
as necessary for the exterior differential to be nilpotent.
We will see later that these charges can be read as ``geometric fluxes'' that, in perfect analogy with the ordinary form fluxes, give origins to the electric and magnetic charges of the effective theory.
These manifolds clearly contain the half-flat manifolds \cite{Chiossi:2002kf}, defined by 
\begin{equation}
{\rm Im} W_1 = {\rm Im} W_2 = W_4 = W_5 = 0.
\label{halfflat} 
\end{equation}
They also include the generalised half flat manifolds, admitting all torsion classes to be non-vanishing, but $W_4 = W_5 = 0$ \cite{deCarlos:2005kh,DAuria:2004tr}.

From the previous discussion we can however argue that this structure can be obtained in a more general setup, where $\widehat J$ and $\widehat \Omega$ do not have vanishing $W_4$ and/or $W_5$, provided the K\"ahler potential for the moduli space is redefined in a proper way.
This is also supported by the comments in \cite{deAlwis:2005tf}, where it is shown that a non-trivial warp factor does not change the superpotential of the effective theory (that is defined through $J$ and $\Omega$), but rather on the K\"ahler potential, which is crucially related to the norm of the structure forms (\ref{KJ}), (\ref{KOmega}).

Coming back to the case of the heterotic strings, we can see how the moduli space for reductions on SU(3) structure manifolds is larger than that of the supersymmetric solutions.
In the case of constant dilaton and warp factor, these are indeed special-hermitian manifolds with $W_1 = W_2 = W_4 = W_5 = 0$ \cite{Cardoso:2002hd}, more constrained than the generic half-flat ones for which $W_4 = W_5 = 0$.
We can then understand how the superpotential of the effective theory can really fix some of the original moduli at the supersymmetric point, in perfect analogy with the IIB case.

\section{Heterotic attractors}\label{sec:attractors}

Let us now come to the realization of the new attractor mechanism, in the case of the common sector of string theory and in particular of the heterotic theory.
We mainly focus on the latter because compactifications on an SU(3) structure manifold give naturally ${\cal N} = 1$ theories in 4 dimensions for the heterotic theory, without the need of orientifolds as in type II.
In this case the role of the sources of negative energy is played by the appearance of the higher-derivative terms in the low-energy lagrangian rather than O-planes.

As a first step let us recall here the basic ingredients.
For the sake of simplicity we focus on the case of trivial warp factor as well as 10-dimensional dilaton, therefore following the example of the previous section.
We then consider compactifications on generalized half-flat manifolds, satisfying the relations (\ref{domi}) and (\ref{dbeta}).
The only form flux that can be turned on is the NS 3-form 
\begin{equation}
H = p^{\Lambda} \alpha_{\Lambda}- q_{\Lambda} \beta^{\Lambda}.
\label{Hflux} 
\end{equation}
Since the H-Bianchi identity is non-trivial for the heterotic theory, we have to be careful with the extra constraints on the 3-forms $\alpha$ and $\beta$ that come from 
\begin{equation}
dH = \tilde \omega^i \left(p^{\Lambda} e_{i \Lambda} - q_{\Lambda}m_i^{\Lambda}\right) = \tilde \omega^i \left(f_i - r_i\right) = \frac{\alpha^{\prime}}{4}\left( {\rm tr} F\wedge F- {\rm tr} R \wedge R\right).
\label{BIJ} 
\end{equation}
We will not look at the effects of the gauge moduli, but only focus on the geometric ones.
Also, we do not look at the specific form of $r_i$ and $f_i$ as they do not affect the attractor mechanism for the definition of the critical points, but keep in mind that extra constraints on the fluxes may arise \cite{deCarlos:2005kh}.
For the standard embedding, or in the case of the common sector of the type I/II theories the right-hand side of (\ref{BIJ}) is vanishing, and then the charges have to fulfill the relation 
\begin{equation}
p^{\Lambda} e_{i \Lambda} - q_{\Lambda}m_i^{\Lambda} = 0.
\label{relcharges} 
\end{equation}

The superpotential is \cite{Cardoso:2003af,Becker:2003gq,Gurrieri:2004dt} 
\begin{equation}
W = \int \left(H + dJ_c\right) \wedge \Omega.
\label{supohet} 
\end{equation}
Using the previous relations, this reads 
\begin{equation}
W = q_{\Lambda} X^{\Lambda}(U) - p^{\Lambda} F_{\Lambda}(U) + T^i \left(m_i^{\Lambda} F_{\Lambda} - X^{\Lambda} e_{i \Lambda}\right), \label{supo2} 
\end{equation}
where $T^i$ are the K\"ahler moduli related to $J_c$ according to (\ref{Jcexp}), and $X^{i}/X^{0} \equiv T^i$.
Introducing the general definition for the ``geometric fluxes'' 
\begin{equation}
F_i \equiv d \omega_i = m_i^{\Lambda} \alpha_{\Lambda}-e_{i \Lambda} \beta^{\Lambda} , \label{geomfl} 
\end{equation}
we can rewrite (\ref{supohet}) as 
\begin{equation}
W = \int \left(H - T^i F_i\right) \wedge \Omega.
\label{supo3} 
\end{equation}

\subsection{One K\"ahler modulus}\label{sub:1mod}

The case of a single size modulus $T^i = T^1 = T$ (or when just one geometric flux is turned on) reduces the previous superpotential to a form very similar to that of type IIB reductions on a Calabi--Yau plus fluxes: 
\begin{eqnarray}
W_{\rm het} & = & \int \left(H - T F\right)\wedge \Omega , \label{Whet1mod} \\[2mm]
W_{IIB} & = & \int \left(F - \tau H\right)\wedge \Omega, \label{WIIBrep} 
\end{eqnarray}
or, using the complex-structure sections:
\begin{eqnarray}
W_{\rm het} & = & \left(e_{\Lambda}^{\rm NS} - T e_{\Lambda}^{\rm geom}\right)X^{\Lambda} - \left(m^{\Lambda}_{\rm NS} - T m^{\Lambda}_{\rm RR}\right) F_{\Lambda}, \label{Whet1mod2} \\[2mm]
W_{IIB} & = & \left(e_{\Lambda}^{\rm RR} - \tau e_{\Lambda}^{\rm NS}\right)X^{\Lambda} - \left(m^{\Lambda}_{\rm RR} - \tau m^{\Lambda}_{\rm NS}\right) F_{\Lambda}.
\label{WIIBrep2} 
\end{eqnarray}
In these expressions, the role of the IIB NS flux is taken by the geometric flux of the heterotic reductions, the IIB RR flux is replaced by the NS 3-form heterotic flux, and the complex dilaton $\tau$ is replaced by the complexified volume modulus $T$.
This clearly follows the pattern of duality relations between type IIB compactifications in the presence of orientifolds and the heterotic theory, through F-theory.
This relation, known for the case of Calabi--Yau compactifications, has recently been studied and extended to the case of non-trivial fluxes in \cite{Becker:2003yv,Becker:2003sh,Becker:2004qh,Becker:2004ii}.
Furthermore, it is clear that, while in type IIB reductions the superpotential (\ref{WIIBrep}) allows all the complex structure moduli to be fixed as well as the dilaton $\tau$, but not the volume, the heterotic potential (\ref{Whet1mod}) fixes all the complex structure moduli and the volume, but not the dilaton.

For this case, where the analogy is clear, we can further see that the minimization of $W_{\rm het}$ gives, for the complexified flux $G_{\rm het} = H -T F$, the same conditions as $W_{\rm IIB}$ imposes on $G_{\rm IIB}$, i.e.~it is restricted to a (2,1) form and primitive for supersymmetric Minkowski vacua and to (2,1)+(0,3) forms for supersymmetric AdS vacua (when neglecting the K\"ahler moduli in type IIB).
We can finally argue that this implies the right vacuum condition for supersymmetric heterotic backgrounds with fluxes \cite{Strominger:1986uh,Cardoso:2002hd}.
Let us see this for the Minkowski vacua.
The internal manifold must be complex as $dJ^{(3,0)} = dJ(0,3) = 0 = d \Omega$.
Moreover, $G^{(1,2)}_{\rm het} = 0$ implies (for the choice $T = i$) that $$H^{(1,2)} + i \overline 
\partial J =0,$$ where we used the fact that $d = 
\partial + \bar 
\partial$ for a complex manifold.
Since $H$ is real we finally obtain 
\begin{equation}
H = i \left( 
\partial - \overline 
\partial\right)J, \label{HJ} 
\end{equation}
which is the condition in \cite{Strominger:1986uh}.
It should be noted that to exhibit non-trivial solutions, the charges cannot be arbitrary, as shown by the previous equation, but the geometrical fluxes must be related to those coming from the 3-form.
This happens because the choice of independent sections $X^{\Lambda}$ imposes that the charges be not independent when trying to achieve the $W = 0$ condition.

It is now also possible to use once more this analogy to extend the new attractor mechanism to this instance of flux compactifications.
Following \cite{Kallosh:2005ax}, we can define generalized symplectic sections 
\begin{equation}
\Pi = \left({\cal V}, -T {\cal V}\right), \label{Pi} 
\end{equation}
where now ${\cal V} = (L^{\Lambda}, M_{\Lambda})$ are only the symplectic sections of the complex structure moduli space.
This doublet of sections couples to the doublet of fluxes 
\begin{equation}
F = (H_3, F_3) \label{Fdoub} 
\end{equation}
to define the central charge 
\begin{equation}
Z = {\rm e}^{K/2}\langle F, \Pi \rangle.
\label{cc} 
\end{equation}
Here $F$ and $\Pi$ are matrices in SL$(2,{\mathbb Z}) \otimes$ Sp$(b_{\rm odd}+1,{\mathbb R})$ and the symplectic pairing $\langle,\rangle$ also contains an SL$(2,{\mathbb Z})$ invariant product of the doublets.
In type IIB this coupling is justified by the SL$(2,{\mathbb Z})$ symmetry of the theory, and by the fact that $\tau$ transforms in the appropriate way.
On the other hand, in the case of the heterotic theory, the justification for such doubling of the symplectic sections is due to the appearance of the K\"ahler modulus $T$.
The pairing (\ref{cc}) of the IIB theory becomes a double symplectic product of the fluxes that transform in both representations of the respective symplectic sections.
In the same way as \cite{Kallosh:2005ax}, we can therefore obtain an algebraic equation for the critical points that is formally the supersymmetric attractor equation 
\begin{equation}
\left( 
\begin{array}{c}
	{\cal Q}_F \\[2mm]
	{\cal Q}_{H} 
\end{array}
\right) = \left( 
\begin{array}{c}
	2 {\rm Re}(\bar Z {\cal V}) \\[2mm]
	2 {\rm Re}(\bar Z T {\cal V}) 
\end{array}
\right)+ \left( 
\begin{array}{c}
	2 {\rm Re}(g^{\alpha \bar \beta}(K_{\overline T})^{-1}\;\overline D_{\overline T}\overline D_{\bar \beta}\bar Z \;D_{\alpha}{\cal V}) \\[2mm]
	2 {\rm Re}(g^{\alpha \bar \beta}(K_{\overline T})^{-1}\;\overline D_{\overline T}\overline D_{\bar \beta}\bar Z \;\bar T D_{\alpha}{\cal V}) 
\end{array}
\right). \label{attractor1} 
\end{equation}
In the IIB case, this formula also follows from the reality of the 4-form flux containing both the RR and NS fluxes in F-theory and its expansion on the basis of 4-forms of the Calabi--Yau 4-fold.
Here we can use the same uplift, but with a different interpretation.
This equation was obtained from the general expansion upon using the supersymmetry condition $D_{\alpha} Z = D_{T} Z = 0$. 
From the same general expansion one could obtain also the non-supersymmetric critical points by imposing the more general conditions deriving from the minimization of the full potential.

Thinking about the expansion in terms of the K\"ahler moduli in (\ref{attractor1}) and looking at the covariantly holomorphic sections, we can see that the charges associated to the geometric fluxes $(e_{i \Lambda}, m_{i}^{\Lambda})$ can be put together with the 3-form flux charges $(m_{0}^{\Lambda},e_{0 \Lambda}) = (p^{\Lambda},q_{\Lambda})$.
Together, they become part of the same symplectic vector in the K\"ahler sector $(m_{I}, e_I)$.
We can therefore argue that the generic charge matrix ${\cal Q}$ is doubly symplectic 
\begin{equation}
{\cal Q} = \left( 
\begin{array}{cc}
	\tilde m^{I\Lambda} & m_{I}^{\Lambda} \\[2mm]
	\tilde e^{I}_{\Lambda} & e_{I \Lambda} 
\end{array}
\right).
\label{Qdef} 
\end{equation}
In the current example, it is clear that only the first column is different from zero as $\tilde e = 0=\tilde m$.
Moreover, for the case of a single K\"ahler modulus, each row in (\ref{Qdef}) transforms generically in Sp(4,${\mathbb R}$), but the lower index $I$ is acted upon only by the SL$(2,{\mathbb R})$ subgroup, as expected from the previous discussion.
We can therefore think about the charges as a matrix to be generically coupled to the product of symplectic sections:
\begin{equation}
{\cal V} = \left( 
\begin{array}{cc}
	L^{I}(T) L^{\Lambda}(U) & M_{I}(T) L^{\Lambda}(U) \\[2mm]
	L^{I}(T) M_{\Lambda}(U) & M_{I}(T) M_{\Lambda}(U) 
\end{array}
\right).
\label{matrsympl} 
\end{equation}

We recall that the ordinary fluxes are quantized, and therefore $e$ and $m$ are generically integers in the appropriate units.
On the other hand, the geometrical fluxes are not usually thought to be quantized; the corresponding charges can therefore have any arbitrary real value.
It is however clear that, as in the previous example, the geometrical fluxes are often related to the ordinary ones by duality relations, so that we expect that also the geometric charges to be quantized.

\subsection{General case}\label{sub:general}

An alternative rewriting of the results of the previous section comes by using the special-K\"ahler properties of the scalar manifold and their formulation in terms of the normalized forms ${\rm e}^{- \hat J}$ and $\widehat \Omega$.
Using the definition (\ref{sec}) for the symplectic sections ${\cal V}$, we can build the corresponding charges 
\begin{equation}
{\cal Q} = \star 1 \otimes H + \widehat \omega \otimes F, \label{Qform} 
\end{equation}
giving a form representation to (\ref{Qdef}) for the case of one volume modulus.
Here $\widetilde \omega$ is such that $\int \omega \wedge \widetilde \omega = 1$, for $J_c = T \omega$.
In this way the central charge (and hence the superpotential $W = {\rm e}^{-K/2}Z$) becomes 
\begin{equation}
Z = \langle {\cal Q}, {\cal V}\rangle, \label{supoQV} 
\end{equation}
where the brackets denote a double Mukai pairing.
A similar proposal, for the generic superpotential coming from type II compactifications on generalized Calabi--Yau manifolds, was put forward in \cite{Berglund:2005dm}, where also matrix charges were considered (see also \cite{Lust:2005bd}).
The result of \cite{Berglund:2005dm} was obtained by computing the non-perturbative contributions to the superpotential, using F-theory uplifts.
Since our definition involves only the perturbative fluxes instead, we can be confident of the fact that this structure will not be spoiled by non-perturbative corrections.
The expansion of the double-symplectic section ${\cal V}$ corresponds to (\ref{matrsympl}) 
\begin{equation}
{\cal V} = \left( 
\begin{array}{cc}
	L^{I}(T) \tilde \omega_{I}\otimes L^{\Lambda}(U) \alpha_{\Lambda} & M_{I}(T)\omega^{I} \otimes L^{\Lambda}(U) \alpha_{\Lambda} \\[2mm]
	L^{I}(T) \tilde \omega_{I} \otimes M_{\Lambda}(U) \beta^{\Lambda} & M_{I}(T) \omega^{I}\otimes M_{\Lambda}(U) \beta^{\Lambda} 
\end{array}
\right), \label{prova} 
\end{equation}
and the corresponding expansion of the charges (\ref{Qform}) on the same basis gives the attractor equation (\ref{attractor1}).

Using this formulation, it is possible to extend these results to the general case of an arbitrary number of K\"ahler moduli.
When there are more active moduli, and therefore more geometrical fluxes, the generic charge matrix becomes 
\begin{equation}
{\cal Q} = \star 1\otimes H + \widetilde \omega^{i} \otimes F_i, \label{Qform2} 
\end{equation}
so that we obtain once more the right central charge from (\ref{supoQV}).

It is again clear that the matrix ${\cal Q}$ does not contain elements in $\Lambda^{2}$ and $\Lambda^{0}$ because of the structure of the generalized half-flat manifolds and of the form-fluxes allowed by the theory, as there is only a 3-form $H$.
We will see that this may change in other theories.
Using this generic expansion in the basis of $\Lambda^{\rm fin}$ we can now write the general new attractor equation for flux vacua of the common sector of string theory.
This reads 
\begin{equation}
\begin{array}{rcl}
	{\cal Q} &=& -2 {\rm Im}\left(\overline Z {\cal V} + g^{\alpha \bar \beta}\; {\rm e}^{-\widehat J_c} \otimes D_{\alpha}\widehat\Omega \; \overline D_{\bar \beta}\overline Z+ g^{i \bar \jmath}\;D_{i}{\rm e}^{-\widehat J_c} \otimes \widehat\Omega\; \overline D_{\bar \jmath}\overline Z \right.\\
	&& \left.+ g^{\alpha \bar \beta} g^{i \bar \jmath}\; D_{i}{\rm e}^{-\widehat J_c} \otimes D_{\alpha}\widehat\Omega\;\overline D_{\bar \jmath}\overline D_{\bar \beta}\overline Z\right) ,
\end{array}
\label{attractorequatio} 
\end{equation}
where the critical point condition for a supersymmetric vacuum $D Z = 0$, or Minkowski $DZ = Z = 0$, or generic non-supersymmetric vacuum $DV = 0$ has to be inserted.
In the special case of supersymmetric Minkowski vacua this simplifies to 
\begin{equation}
{\cal Q} = -2 \, {\rm Im}\left( g^{\alpha \bar \beta} g^{i \bar \jmath}\; D_{i}{\rm e}^{-\widehat J_c} \otimes D_{\alpha}\widehat\Omega\;\overline D_{\bar \jmath}\overline D_{\bar \beta}\overline Z\right),\label{Mink} 
\end{equation}
which implies once more that the geometrical fluxes contain only a primitive $dJ^{(2,1)+(1,2)}$, therefore giving as attracting manifolds the special-hermitian ones.
This can be seen from the fact that $D_{i}{\rm e}^{-\widehat J_c} \otimes D_{\alpha}\widehat\Omega$ selects from the charges the elements in $D_i M_j \tilde \omega^j \otimes \chi_{\alpha}^{(2,1)}$ and $D_i M_0 \star 1 \otimes \chi_{\alpha}^{(2,1)}$.
It should be recalled that only for vanishing cosmological constant are these real points in the landscape of flux vacua for the common sector.
The full K\"ahler potential depends also on the dilaton $S$, whereas the superpotential does not.
This means that the conditions for a critical point with respect to $S$ give $D_{S} W = K_{S} W = 0$.
The same is true also for the type IIB case, where the role of the dilaton is taken by the volume modulus.
It is indeed known that, without non-perturbative effects, the IIB  potential is of the no-scale form.
In any case, understanding the general condition may be of some use for KKLT-like scenarios and as preparation for the IIA case, where this is no longer a problem and the superpotential can depend on all the moduli.
We notice again that only some of the charges were non-trivial so far.
We will see that, without non-perturbative contributions, we cannot produce all non-trivial entries.

\section{Type II non-K\"ahler attractors}\label{sec:typeII}

Now that we have seen how the new attractor mechanism works for the case of the heterotic string, we give some comments on how this should be further extended to the case of type II compactifications on non-K\"ahler manifolds with O-planes.
As we have discussed in the introduction, the justification for this type of backgrounds is given by the extension of mirror symmetry to the flux case or, more generally, by using T-duality on flux backgrounds.

When type II theories are compactified on an SU(3) structure manifold, we get an ${\cal N} = 2$ effective theory \cite{Grana:2005ny}.
In order to reduce to an ${\cal N} = 1$ lagrangian we must add orientifold projections.
This, however, means that the spectrum corresponding to this compactification is truncated and may affect the special-K\"ahler structure of the moduli space.
In the case of type IIA with orientifolds, we can see that there is a relic of the original special-K\"ahler structure in the size deformation part of the moduli space as the scalar fields in this sector are described by the components of $J_c = B + i J$ that survive the projections.
On the other hand, where the complex structure moduli are concerned, the orientifold projections remove the real parts of the complex moduli.
However, these are replaced by the surviving moduli coming from the RR 3-form $C$.
This sector of the moduli space can now be described by the new ``holomorphic form'' 
\begin{equation}
\Omega_c = C_3 + i {\rm e}^{-\Phi} {\rm Im} \Omega.
\label{Omc} 
\end{equation}
The K\"ahler potential of the moduli space is therefore still given by expressions that formally resemble the ones in (\ref{KJ}) and (\ref{KOmega}), but now replaced by the new sections 
\begin{equation}
K_{J} = -\log i \langle {\rm e}^{- J_c}, {\rm e}^{-\overline J_c}\rangle, \quad K_{\Omega} = - \log i \langle \Omega_c, \overline \Omega_{c}\rangle.
\label{newpot} 
\end{equation}
In addition, we can now have non-trivial RR 2- and 4-form fluxes.
These fluxes can be expanded in the same basis as before 
\begin{eqnarray}
g_2 & = & \tilde e_{\rm RR}^i \omega_i, \label{g2} \\[2mm]
f_4 & = & e^{\rm RR}_i \tilde \omega^i, \label{f4} 
\end{eqnarray}
where the charges are constrained by the closure of the Bianchi identities $dg_2 =0$ and $df_4 = H \wedge f_2 = 0$.
These constraints read \cite{Grana:2005ny}
\begin{equation}
\tilde e_{\rm RR}^i e_{i \Lambda} = 0, \qquad \tilde e^{i}_{\rm RR} m_i^{\Lambda} = 0.
\label{label3} 
\end{equation}
The equations of motion give further $e_{i}^{\rm RR} g^{ij}(U) e_{j \Lambda} = 0 = e_{i}^{\rm RR} g^{ij}(U) m_j^{\Lambda}$.
The charges in (\ref{g2})--(\ref{f4}) are electric-magnetic duals under symplectic rotations of the K\"ahler deformations part, labelled by the indices $I =(0,i)$, and therefore constitute part of a symplectic vector $(\tilde e^I,e_I)$.
They have the same properties for the electric-magnetic transformations of the complex-structure deformations and this is why we call them both with the same letter $e$.
We will see that this is also consistent with our previous definition (\ref{Qdef}).

The superpotential of the effective theory has been obtained in various ways \cite{Gurrieri:2002wz,Grimm:2004ua,Grana:2005ny,Villadoro:2005cu,DallAgata:2005fm}, and it reads 
\begin{equation}
W = \int J_c \wedge d \Omega_c - \int H \wedge \Omega_c + \frac12 \int J_c \wedge J_c \wedge g_{2} + \int f_{4} \wedge J_c, \label{supoIIA} 
\end{equation}
where, once the first term is integrated by parts, we recover the superpotential of the common sector from the first two terms.

At this stage we can extend the arguments of the previous section by enlarging the charge matrix 
\begin{equation}
{\cal Q} = {\cal Q}^{\rm NS} + {\cal Q}^{\rm RR}, \label{Qgen} 
\end{equation}
where ${\cal Q}^{\rm NS} $ is given by (\ref{Qform2}), and ${\cal Q}^{\rm RR}$ can now be defined as\footnote{The vector of RR charges can be completed to a full symplectic vector in the $I$ indices by the 0- and 6-form fluxes that appear for instance in massive type IIA. 
These are expanded as $g_0 = \tilde e^{0}_{RR}$, $g_6 = e_0^{RR} \star 1$.} 
\begin{equation}
{\cal Q}^{\rm RR} = f_4 \otimes \Xi + g_2 \otimes \Xi = \tilde e^{i}_{\rm RR} \omega_i \otimes \Xi+ e_i^{\rm RR} \tilde \omega^i \otimes \Xi, \label{QIIA} 
\end{equation}
where $\Xi = {\rm e}^{K_{\Omega}/2} i (\widehat \Omega - \overline{ \widehat \Omega })$ is a real 3-form chosen so that its contribution to the superpotential is trivial.
However, we expect that also $(\tilde e^I_{RR}, e^{RR}_I)$ become part of symplectic vector in $\Lambda$.
For this to be the case, we have to choose $\Xi = \beta^0$, so that the central charge dependence on the complex structure moduli is only through $L^0$, that can be fixed to be $L^0 = {\rm e}^{K_\Omega/2}$, in a way compatible with the superpotential presented in \cite{Grana:2005ny}.
Once more, the central charge is obtained by the symplectic contraction $$ Z = \langle {\cal Q}, {\cal V}\rangle.
$$ In components, the RR part of the superpotential reads 
\begin{equation}
W = X^{i}(T) e_{i}^{\rm RR} -  \tilde e^{i}_{\rm RR} F_{i}(T).
\label{WIIA comp} 
\end{equation}
Finally, we can expand again the charge matrix ${\cal Q}$ in terms of the projections on the various sectors of $\Lambda^{even} \otimes \Lambda^{odd}$ as before, although now the charge matrix has additional non-trivial entries 
\begin{equation}
{\cal Q} = \left( 
\begin{array}{cc}
	 0 & m_{I}^{\Lambda} \\
	 (\tilde e^I_0,0) & e_{I \Lambda} 
\end{array}
\right), \label{QIIa} 
\end{equation}
and the $e_{I \Lambda}$ now include both the NS and RR charges.
It has been noted that the conditions on the fluxes, here following from the differential algebra, have an interpretation in the effective field theory in terms of gauging conditions \cite{DAuria:2004tr,DAuria:2004wd,Grana:2005ny}.
It would be interesting to spell the exact conditions on the charge matrix (as in equations (\ref{const}), (\ref{relcharges}) and (\ref{label3}) presented above) independently from the structure of the flux compactifications presented here.

In an analogous fashion we can argue for the extension of this mechanism to the case of IIB compactifications on non-K\"ahler manifolds.
For this case we have to add 
\begin{equation}
{\cal Q}_{\rm RR}^{IIB} =\star 1 \otimes F_{\rm RR}.
\label{QIIB} 
\end{equation}
In this case, as is evident from the additional SL$(2,{\mathbb Z})$ symmetry of the theory and from the fact that the complex dilaton appears explicitly in the superpotential, the ${\cal Q}$ and ${\cal V}$ matrices must be further enlarged as in \cite{Kallosh:2005ax}, where the new sections form explicit doublets of this symmetry.
However, once this technical trick is implemented, the attractor equations are obtained in the same way as presented in the previous sections.

\bigskip 

\bigskip

\noindent {\bf Acknowledgments}

\medskip

\noindent I would like to thank R.~Kallosh, S.~Ferrara and A.~Lukas for valuable comments and discussions.

%%%%%%%%%%%%%%%%% UNCOMMENT THIS PART FOR BIBTEX %%%%%%%%%%%%%%%%
% \bibliography{totArXiv} 

\begingroup 
\begin{thebibliography}
{10}

\bibitem{Ferrara:1995ih} S.~Ferrara, R.~Kallosh and A.~Strominger, {\it N=2 extremal black holes}, {\em Phys.
Rev.} {\bf D52} (1995) 5412--5416 [\href{http://arXiv.org/abs/hep-th/9508072}{{\tt hep-th/9508072}}].

%%CITATION = HEP-TH 9508072;%%
\bibitem{Ferrara:1997tw} S.~Ferrara, G.~W.
Gibbons and R.~Kallosh, {\it Black holes and critical points in moduli space}, {\em Nucl.
Phys.} {\bf B500} (1997) 75--93 [\href{http://arXiv.org/abs/hep-th/9702103}{{\tt hep-th/9702103}}].

%%CITATION = HEP-TH 9702103;%%
\bibitem{Andrianopoli:1996ve} L.~Andrianopoli, R.~D'Auria and S.~Ferrara, {\it U-duality and central charges in various dimensions revisited}, {\em Int.
J.
Mod.
Phys.} {\bf A13} (1998) 431--490 [\href{http://arXiv.org/abs/hep-th/9612105}{{\tt hep-th/9612105}}].

%%CITATION = HEP-TH 9612105;%%
\bibitem{Ferrara:1996dd} S.~Ferrara and R.~Kallosh, {\it Supersymmetry and attractors}, {\em Phys.
Rev.} {\bf D54} (1996) 1514--1524 [\href{http://arXiv.org/abs/hep-th/9602136}{{\tt hep-th/9602136}}].

%%CITATION = HEP-TH 9602136;%%
\bibitem{Ceresole:1995ca} A.~Ceresole, R.~D'Auria and S.~Ferrara, {\it The symplectic structure of n=2 supergravity and its central extension}, {\em Nucl.
Phys.
Proc.
Suppl.} {\bf 46} (1996) 67--74 [\href{http://arXiv.org/abs/hep-th/9509160}{{\tt hep-th/9509160}}].

%%CITATION = HEP-TH 9509160;%%
\bibitem{Ceresole:1995jg} A.~Ceresole, R.~D'Auria, S.~Ferrara and A.~Van~Proeyen, {\it Duality transformations in supersymmetric Yang--Mills theories coupled to supergravity}, {\em Nucl.
Phys.} {\bf B444} (1995) 92--124 [\href{http://arXiv.org/abs/hep-th/9502072}{{\tt hep-th/9502072}}].

%%CITATION = HEP-TH 9502072;%%
\bibitem{Strominger:1996kf} A.~Strominger, {\it Macroscopic entropy of $n=2$ extremal black holes}, {\em Phys.
Lett.} {\bf B383} (1996) 39--43 [\href{http://arXiv.org/abs/hep-th/9602111}{{\tt hep-th/9602111}}].

%%CITATION = HEP-TH 9602111;%%
\bibitem{Moore:1998pn} G.~W.
Moore, {\it Arithmetic and attractors}, \href{http://arXiv.org/abs/hep-th/9807087}{{\tt hep-th/9807087}}.

%%CITATION = HEP-TH 9807087;%%
\bibitem{Kallosh:2005bj} R.~Kallosh, {\it Flux vacua as supersymmetric attractors}, \href{http://arXiv.org/abs/hep-th/0509112}{{\tt hep-th/0509112}}.

%%CITATION = HEP-TH 0509112;%%
\bibitem{Kallosh:2005ax} R.~Kallosh, {\it New attractors}, \href{http://arXiv.org/abs/hep-th/0510024}{{\tt hep-th/0510024}}.

%%CITATION = HEP-TH 0602005;%%
\bibitem{Kallosh:2006} R.~Kallosh, N.~Sivanandam and M.~Soroush, {\it The non-BPS black hole attractor equation}, \href{http://arXiv.org/abs/hep-th/0602005}{{\tt hep-th/0602005}}.


%%CITATION = HEP-TH 0510024;%%
\bibitem{Strominger:1986uh} A.~Strominger, {\it Superstrings with torsion}, {\em Nucl.
Phys.} {\bf B274} (1986) 253.

%%CITATION = NUPHA,B274,253;%%
\bibitem{Kachru:2002sk} S.~Kachru, M.~B.
Schulz, P.~K.
Tripathy and S.~P.
Trivedi, {\it New supersymmetric string compactifications}, {\em JHEP} {\bf 03} (2003) 061 [\href{http://arXiv.org/abs/hep-th/0211182}{{\tt hep-th/0211182}}].

%%CITATION = HEP-TH 0211182;%%
\bibitem{Gurrieri:2002wz} S.~Gurrieri, J.~Louis, A.~Micu and D.~Waldram, {\it Mirror symmetry in generalized Calabi--Yau compactifications}, {\em Nucl.
Phys.} {\bf B654} (2003) 61--113 [\href{http://arXiv.org/abs/hep-th/0211102}{{\tt hep-th/0211102}}].

%%CITATION = HEP-TH 0211102;%%
\bibitem{Grana:2005ny} M.~Gra\~na, J.~Louis and D.~Waldram, {\it Hitchin functionals in N = 2 supergravity}, \href{http://arXiv.org/abs/hep-th/0505264}{{\tt hep-th/0505264}}.

%%CITATION = HEP-TH 0505264;%%


\bibitem{Cardoso:2002hd} G.~L.
Cardoso et.~al., {\it Non-Kaehler string backgrounds and their five torsion classes}, {\em Nucl.
Phys.} {\bf B652} (2003) 5--34 [\href{http://arXiv.org/abs/hep-th/0211118}{{\tt hep-th/0211118}}].

%%CITATION = HEP-TH 0211118;%%

\bibitem{House:2005yc}
  T.~House and E.~Palti,
  {\it Effective action of (massive) IIA on manifolds with SU(3) structure},
  Phys.\ Rev.\ D {\bf 72} (2005) 026004
  [arXiv:hep-th/0505177].
  %%CITATION = HEP-TH 0505177;%%


\bibitem{Candelas:1990pi} P.~Candelas and X.~de~la Ossa, {\it Moduli space of Calabi--Yau manifolds}, {\em Nucl.
Phys.} {\bf B355} (1991) 455--481.


\bibitem{deCarlos:2005kh} B.~de~Carlos, S.~Gurrieri, A.~Lukas and A.~Micu, {\it Moduli stabilisation in heterotic string compactifications}, \href{http://arXiv.org/abs/hep-th/0507173}{{\tt hep-th/0507173}}.

%%CITATION = HEP-TH 0507173;%%
\bibitem{DAuria:2004tr} R.~D'Auria, S.~Ferrara, M.~Trigiante and S.~Vaul\'a, {\it Gauging the Heisenberg algebra of special quaternionic manifolds}, {\em Phys.
Lett.} {\bf B610} (2005) 147--151 [\href{http://arXiv.org/abs/hep-th/0410290}{{\tt hep-th/0410290}}].

\bibitem{Chiossi:2002kf} S.~Chiossi and S.~Salamon, {\it The intrinsic torsion of {SU(3)} and {$G_2$} structures}, in {\em Differential geometry, Valencia, 2001}, pp.~115--133.
\newblock World Sci.
Publishing, River Edge, NJ, 2002.
\newblock \href{http://arXiv.org/abs/math.DG/0202282}{{\tt math.DG/0202282}}.


%%CITATION = HEP-TH 0412063;%%
\bibitem{deAlwis:2005tf} S.~P.
de~Alwis, {\it Effective potentials for light moduli}, \href{http://arXiv.org/abs/hep-th/0506266}{{\tt hep-th/0506266}}.

%%CITATION = HEP-TH 0506266;%%


\bibitem{Cardoso:2003af} G.~L.
Cardoso, G.~Curio, G.~Dall'Agata and D.~L\"ust, {\it BPS action and superpotential for heterotic string compactifications with fluxes}, {\em JHEP} {\bf 10} (2003) 004 [\href{http://arXiv.org/abs/hep-th/0306088}{{\tt hep-th/0306088}}].

%%CITATION = HEP-TH 0306088;%%
\bibitem{Becker:2003gq} K.~Becker, M.~Becker, K.~Dasgupta and S.~Prokushkin, {\it Properties of heterotic vacua from superpotentials}, {\em Nucl.
Phys.} {\bf B666} (2003) 144--174 [\href{http://arXiv.org/abs/hep-th/0304001}{{\tt hep-th/0304001}}].

%%CITATION = HEP-TH 0304001;%%
\bibitem{Gurrieri:2004dt} S.~Gurrieri, A.~Lukas and A.~Micu, {\it Heterotic on half-flat}, {\em Phys.
Rev.} {\bf D70} (2004) 126009 [\href{http://arXiv.org/abs/hep-th/0408121}{{\tt hep-th/0408121}}].

%%CITATION = HEP-TH 0408121;%%
\bibitem{Becker:2003yv} K.~Becker, M.~Becker, K.~Dasgupta and P.~S.
Green, {\it Compactifications of heterotic theory on non-Kaehler complex manifolds.
I}, {\em JHEP} {\bf 04} (2003) 007 [\href{http://arXiv.org/abs/hep-th/0301161}{{\tt hep-th/0301161}}].

%%CITATION = HEP-TH 0301161;%%
\bibitem{Becker:2003sh} K.~Becker, M.~Becker, P.~S.
Green, K.~Dasgupta and E.~Sharpe, {\it Compactifications of heterotic strings on non-Kaehler complex manifolds.
II}, {\em Nucl.
Phys.} {\bf B678} (2004) 19--100 [\href{http://arXiv.org/abs/hep-th/0310058}{{\tt hep-th/0310058}}].

%%CITATION = HEP-TH 0310058;%%
\bibitem{Becker:2004qh} M.~Becker, K.~Dasgupta, A.~Knauf and R.~Tatar, {\it Geometric transitions, flops and non-Kaehler manifolds.
I}, {\em Nucl.
Phys.} {\bf B702} (2004) 207--268 [\href{http://arXiv.org/abs/hep-th/0403288}{{\tt hep-th/0403288}}].

%%CITATION = HEP-TH 0403288;%%
\bibitem{Becker:2004ii} K.~Becker, M.~Becker, K.~Dasgupta and R.~Tatar, {\it Geometric transitions, non-Kaehler geometries and string vacua}, {\em Int.
J.
Mod.
Phys.} {\bf A20} (2005) 3442--3448 [\href{http://arXiv.org/abs/hep-th/0411039}{{\tt hep-th/0411039}}].

%%CITATION = HEP-TH 0411039;%%

\bibitem{Berglund:2005dm} P.~Berglund and P.~Mayr, {\it Non-perturbative superpotentials in F-theory and string duality}, \href{http://arXiv.org/abs/hep-th/0504058}{{\tt hep-th/0504058}}.

%%CITATION = HEP-TH 0504058;%%
\bibitem{Lust:2005bd} D.~L\"ust, P.~Mayr, S.~Reffert and S.~Stieberger, {\it F-theory flux, destabilization of orientifolds and soft terms on D7-branes}, \href{http://arXiv.org/abs/hep-th/0501139}{{\tt hep-th/0501139}}.

%%CITATION = HEP-TH 0501139;%%
\bibitem{Grimm:2004ua} T.~W.
Grimm and J.~Louis, {\it The effective action of type IIA Calabi--Yau orientifolds}, {\em Nucl.
Phys.} {\bf B718} (2005) 153--202 [\href{http://arXiv.org/abs/hep-th/0412277}{{\tt hep-th/0412277}}].

%%CITATION = HEP-TH 0412277;%%
\bibitem{Villadoro:2005cu} G.~Villadoro and F.~Zwirner, {\it N = 1 effective potential from dual type-IIA D6/O6 orientifolds with general fluxes}, {\em JHEP} {\bf 06} (2005) 047 [\href{http://arXiv.org/abs/hep-th/0503169}{{\tt hep-th/0503169}}].

%%CITATION = HEP-TH 0503169;%%
\bibitem{DallAgata:2005fm} G.~Dall'Agata and N.~Prezas, {\it Scherk--Schwarz reduction of M-theory on $G_2$-manifolds with fluxes}, \href{http://arXiv.org/abs/hep-th/0509052}{{\tt hep-th/0509052}}.

%%CITATION = HEP-TH 0509052;%%


%%CITATION = HEP-TH 0410290;%%
\bibitem{DAuria:2004wd} R.~D'Auria, S.~Ferrara, M.~Trigiante and S.~Vaul\'a, {\it Scalar potential for the gauged Heisenberg algebra and a non-polynomial antisymmetric tensor theory}, {\em Phys.
Lett.} {\bf B610} (2005) 270--276 [\href{http://arXiv.org/abs/hep-th/0412063}{{\tt hep-th/0412063}}].


\end{thebibliography}

\endgroup
% \bibliographystyle{JHEPmod}
% %%%%%%%%%%% CUT OUT THE REST IF YOU USE BIBTEX
\providecommand{\href}[2]{#2} 
\end{document}